\documentclass{camera}
\input{epsf}

\def\ltap{\raisebox{-.6ex}{\rlap{$\,\sim\,$}} \raisebox{.4ex}{$\,<\,$}} 
\def\gtap{\raisebox{-.6ex}{\rlap{$\,\sim\,$}} \raisebox{.4ex}{$\,>\,$}}

\newcommand\as{\alpha_{\mathrm{S}}}

\def\beq{\begin{equation}} 
\def\eeq{\end{equation}} 
\def\beeq{\begin{eqnarray}} 
\def\eeeq{\end{eqnarray}} 
 
\def\to{\rightarrow}

\def\b0{b_0}

\begin{document}
\title{$Q_T$-resummation in \\ Higgs boson production \\ at the LHC~\footnote{Talk 
given in italian at Incontri sulla Fisica delle Alte Energie, Lecce, Italy, 23-26 April 2003.}}
\author{G.~Bozzi}
\organization{Dipartimento di Fisica, Universit\`a di Firenze and INFN,\\
I-50019 Sesto Fiorentino, Florence, Italy}
\maketitle
\addtolength{\parindent}{-0.5cm}
\rule{12.6cm}{1pt}
\addtolength{\parindent}{0.5cm}
{\bf Abstract}

When considering the transverse momentum distribution ($q_T$) of the Higgs boson 
production it is necessary to separate the small $q_T$ region ($q_T \ll M_H$) from 
the medium and large ($q_T \gtap M_H$) one, the former being treated by means of
resummation techniques of logaritmhically-enhanced contributions and the latter by 
fixed-order perturbation theory. Then these two approaches have to be consistently matched 
to avoid double-counting in the intermediate $q_T$ region.
Here soft gluon resummation is implemented up to NNLL order and the matching 
to the corresponding NLO perturbative result is performed. Numerical results are shown for the LHC.
The main features of the differential distribution turn out to be quite stable with respect
to perturbative uncertainties.\\ 
\rule{12.6cm}{1pt}
\break
An accurate theoretical prediction of the transverse-momentum ($q_T$) distribution
of the Higgs boson at the LHC can be important to enhance the 
statistical significance of the signal over the background and to improve strategies 
for the extraction of the signal [\ref{atlascms}]. In what follows we consider the 
most relevant production mechanism: the gluon initiated process via a top-quark loop. 

It is convenient to consider and treat separately
the large-$q_T$ and small-$q_T$ regions of the spectrum.
Roughly speaking, the large-$q_T$ region is identified by the condition $q_T \gtap M_H$.
In this region, the perturbative series is controlled by a small expansion
parameter, $\as(M_H^2)$, and calculations based on the truncation of the series
at a fixed-order in $\as$ are theoretically justified and reliable.
The LO calculation ${\cal O}(\as^3)$ was reported in Ref.~[\ref{Ellis:1987xu}];
it shows that the large-$M_t$ 
approximation (the limit of an infinitely-heavy top quark)works well as long
as both $M_H$ and $q_T$ are smaller than $M_t$.
In the framework of this approximation, the NLO QCD corrections
were computed first numerically [\ref{deFlorian:1999zd}]
and later analytically [\ref{Ravindran:2002dc}, \ref{Glosser:2002gm}].

In the small-$q_T$ region ($q_T\ll M_H$), where the bulk of events is produced,
the convergence of the fixed-order expansion is spoiled, since
the coefficients of the perturbative series in $\as(M_H^2)$ are enhanced
by powers of large logarithmic terms, $\ln^m (M_H^2/q_T^2)$. To obtain
reliable perturbative predictions, these terms have 
to be systematically resummed to all orders in $\as$ [\ref{Dokshitzer:hw},
see also the list of references in Sect.~5 of
Ref.~\ref{Catani:2000jh}].
In the case of the Higgs boson, resummation has been explicitly worked out at
leading logarithmic (LL), next-to-leading logarithmic (NLL) 
[\ref{Catani:vd}, \ref{Kauffman:cx}]
and next-to-next-to-leading logarithmic (NNLL) [\ref{deFlorian:2000pr}] level.
The fixed-order and resummed approaches have then
to be consistently matched at intermediate values of $q_T$, 
so as to avoid the introduction of ad-hoc 
boundaries between the large-$q_T$ and small-$q_T$ regions.

In this work the formalism described in Ref.~[\ref{Catani:2000vq}] is used to compute  
the Higgs boson $q_T$ distribution at the LHC. In particular, it includes the
most advanced perturbative information that is available at present:
NNLL resummation at small $q_T$ and NLO calculations at large $q_T$.
As the matching procedure could introduce higher order corrections in the intermediate
$q_T$ region, it proves useful to put a constrain on the integral over $q_T$ of the differential
distribution, which should reproduce the total cross section result (known at NLO 
[\ref{Dawson:1991zj}] and NNLO [\ref{NNLOtotal}]).
More details and formulas can be found in Ref.~[\ref{Ours}].
Other recent phenomenological predictions can be found in [\ref{recent}].

In the following, quantitative results at NLL+LO and NNLL+NLO accuracy are presented. 
At NLL+LO accuracy the NLL resummed result is matched with the LO perturbative result,
while at NNLL+NLO accuracy the NNLL resummed result is matched with the NLO perturbative result. 
As for the evaluation of the fixed order results, the Monte Carlo program 
of Ref.~[\ref{deFlorian:1999zd}] has been used.
The numerical results are obtained by choosing $M_H=125$~GeV and using 
the MRST2002 set of parton distributions [\ref{Martin:2003es}].
They slightly differ from those presented in [\ref{Ours}], 
where we used the MRST2001 set [\ref{Martin:2001es}].
At NLL+LO, LO parton densities and 
1-loop $\as$ have been used, whereas NLO parton densities 
and 2-loop $\as$ for the NNLL+NLO matching.

\begin{figure}[htb]
\begin{center}
\begin{tabular}{c}
\epsfxsize=9truecm
\epsffile{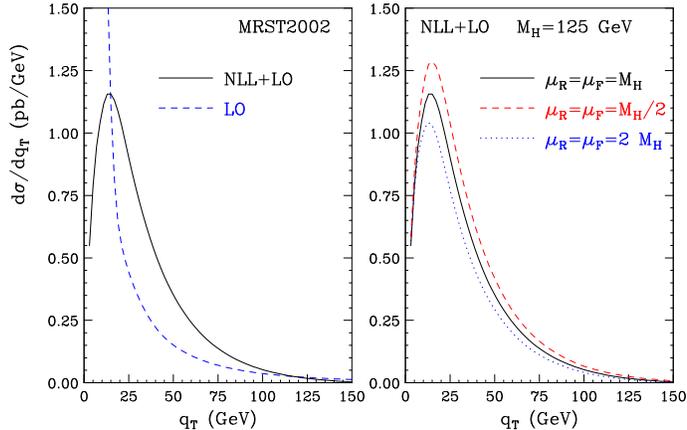}\\
\end{tabular}
\end{center}
\caption{\label{fig1}
{\em 
LHC results at NLL+LO accuracy.}}
\end{figure}

The NLL+LO results at the LHC are shown in Fig.~\ref{fig1}.
In the left-hand side, the full NLL+LO result (solid line)
is compared with the LO one (dashed line)
at the default scales $\mu_F=\mu_R=M_H$.
We see that the LO calculation diverges to $+\infty$ as $q_T\to 0$. 
The effect of the resummation is relevant below $q_T\sim 100$~GeV.
In the right-hand side we show the NLL+LO band that is obtained
by varying $\mu_F=\mu_R$ between $1/2 M_H$ and $2M_H$.
The scale dependence increases from about $\pm 10\%$ at the peak
to about $\pm 20\%$ at $q_T=100$~GeV.

\begin{figure}[htb]
\begin{center}
\begin{tabular}{c}
\epsfxsize=9truecm
\epsffile{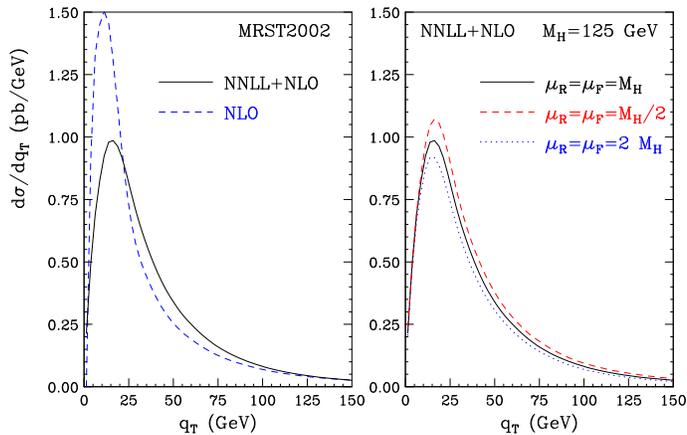}\\
\end{tabular}
\end{center}
\caption{\label{fig2}
{\em 
LHC results at NNLL+NLO accuracy. }}
\end{figure}

The NNLL+NLO results at the LHC are shown in Fig.~\ref{fig2}.
In the left-hand side, the full result (solid line)
is compared with the NLO one (dashed line) at the
default scales $\mu_F=\mu_R=M_H$.
The NLO result diverges to $-\infty$ as $q_T\to 0$ and, at small values of 
$q_T$, it has an unphysical peak (the top of the peak is close to the vertical
scale of the plot) which is produced by the numerical compensation of negative
leading logarithmic and positive subleading logarithmic contributions.
It is interesting to compare the LO and NLL+LO curves in Fig.~\ref{fig1}
and the NLO curve in Fig.~\ref{fig2}. At $q_T \sim 50$~GeV, the 
$q_T$ distribution sizeably increases when going from LO to NLO and from NLO
to NLL+LO. This implies that in the intermediate-$q_T$ region there are
important contributions that have to be resummed to all orders rather than
simply evaluated at the next perturbative order.
The $q_T$ distribution is (moderately) harder at NNLL+NLO than at NLL+LO accuracy.
The height of the NNLL peak is a bit lower than the NLL one. This is mainly due to 
the fact that the total NNLO cross section (computed with NLO parton densities 
and 2-loop $\as$), which fixes the value of the $q_T$ integral of our resummed result,
is slightly smaller than the NLO one, whereas the high-$q_T$ tail is higher at NNLL order,
thus leading to a reduction of the cross section at small $q_T$.
The resummation effect starts to be visible below $q_T\sim 100$~GeV, and 
it increases the NLO result by about $40\%$ at $q_T=50$~GeV.
The right-hand side of Fig.~\ref{fig2} shows the scale dependence computed as
in Fig.~\ref{fig1}. The scale dependence is now about $\pm 8\%$ at the peak
and increases to $\pm 20\%$ at $q_T=100$~GeV.
Comparing Figs.~1 and 2, we see that the NNLL+NLO band is smaller 
than the NLL+LO one and overlaps with the latter at $q_T \ltap 100$~GeV.
This suggests a good convergence of the resummed perturbative expansion. 

\section*{References}
\begin{enumerate}

\item \label{atlascms}
CMS Coll., {\it Technical Proposal}, report CERN/LHCC/94-38 (1994);
ATLAS Coll., {\it ATLAS Detector and Physics Performance: Technical Design
Report}, Vol. 2, report CERN/LHCC/99-15 (1999).
\\[-0.7cm]
\item \label{Ellis:1987xu}
R.~K.~Ellis, I.~Hinchliffe, M.~Soldate and J.~J.~van der Bij,
Nucl.\ Phys.\ B {\bf 297} (1988) 221;
U.~Baur and E.~W.~Glover,
Nucl.\ Phys.\ B {\bf 339} (1990) 38.
\\[-0.7cm]
\item \label{deFlorian:1999zd}
D.~de Florian, M.~Grazzini and Z.~Kunszt,
Phys.\ Rev.\ Lett.\  {\bf 82} (1999) 5209.
\\[-0.7cm]
\item \label{Ravindran:2002dc}
V.~Ravindran, J.~Smith and W.~L.~Van Neerven,
Nucl.\ Phys.\ B {\bf 634} (2002) 247.
\\[-0.7cm]
\item \label{Glosser:2002gm}
C.~J.~Glosser and C.~R.~Schmidt,
JHEP {\bf 0212} (2002) 016.
\\[-0.7cm]
\item \label{Dokshitzer:hw}
G.~Parisi and R.~Petronzio,
Nucl.\ Phys.\ B {\bf 154} (1979) 427;
Y.~L.~Dokshitzer, D.~Diakonov and S.~I.~Troian,
Phys.\ Rep.\  {\bf 58} (1980) 269;
J.~C.~Collins, D.~E.~Soper and G.~Sterman,
Nucl.\ Phys.\ B {\bf 250} (1985) 199.
\\[-0.7cm]
\item \label{Catani:2000jh}
S.~Catani et al.,
hep-ph/0005025, in the Proceedings of the CERN Workshop on {\it Standard Model
Physics (and more) at the LHC}, eds. G.~Altarelli and M.L.~Mangano
(CERN 2000-04, Geneva, 2000), p.~1.
\\[-0.7cm]
\item \label{Catani:vd}
S.~Catani, E.~D'Emilio and L.~Trentadue,
Phys.\ Lett.\ B {\bf 211} (1988) 335.
\\[-0.7cm]
\item \label{Kauffman:cx}
R.~P.~Kauffman,
Phys.\ Rev.\ D {\bf 45} (1992) 1512.
\\[-0.7cm]
\item \label{deFlorian:2000pr}
D.~de Florian and M.~Grazzini,
Phys.\ Rev.\ Lett.\ {\bf 85} (2000) 4678,
Nucl.\ Phys.\ B {\bf 616} (2001) 247.
\\[-0.7cm]
\item \label{Catani:2000vq}
S.~Catani, D.~de Florian and M.~Grazzini,
Nucl.\ Phys.\ B {\bf 596} (2001) 299.
\\[-0.7cm]
\item \label{Dawson:1991zj}
S.~Dawson,
Nucl.\ Phys.\ B {\bf 359} (1991) 283;
A.~Djouadi, M.~Spira and P.~M.~Zerwas,
Phys.\ Lett.\ B {\bf 264} (1991) 440;
M.~Spira, A.~Djouadi, D.~Graudenz and P.~M.~Zerwas,
Nucl.\ Phys.\ B {\bf 453} (1995) 17.
\\[-0.7cm]
\item \label{NNLOtotal}
S.~Catani, D.~de Florian and M.~Grazzini,
JHEP {\bf 0105} (2001) 025;
R.~V.~Harlander and W.~B.~Kilgore,
Phys.\ Rev.\ D {\bf 64} (2001) 013015,
Phys.\ Rev.\ Lett.\  {\bf 88} (2002) 201801;
C.~Anastasiou and K.~Melnikov,
Nucl.\ Phys.\ B {\bf 646} (2002) 220;
V.~Ravindran, J.~Smith, W.~L.~van Neerven,
Nucl.\ Phys.\ B {\bf 665} (2003) 325.
\\[-0.7cm]
\item \label{Ours}
G. ~Bozzi, S.~Catani, D. ~de ~Florian, M.~Grazzini,
Phys.\ Lett.\  {\bf B564} (2003) 65.
\\[-0.7cm]
\item \label{recent}
C.~Balazs and C.~P.~Yuan,
Phys.\ Lett.\ B {\bf 478} (2000) 192;
E.~L.~Berger and J.~w.~Qiu,
Phys.\ Rev.\ D {\bf 67} (2003) 034026;
A.~Kulesza, G.~Sterman, W.~Vogelsang,
preprint BNL-HET-03/20 [hep-ph/0309264].
\\[-0.7cm]
\item \label{Martin:2001es}
A.~D.~Martin, R.~G.~Roberts, W.~J.~Stirling and R.~S.~Thorne,
Eur.\ Phys.\ J.\ C {\bf 23} (2002) 73.
\\[-0.7cm]
\item \label{Martin:2003es}
A.~D.~Martin, R.~G.~Roberts, W.~J.~Stirling and R.~S.~Thorne,
Eur.\ Phys.\ J.\ C {\bf 28} (2003) 455.

\end{enumerate}

\end{document}